\documentclass{llncs}
\usepackage{hyperref}
\usepackage[pdftex]{graphicx}
\usepackage[pdftex]{color}
\usepackage{lmodern}
\usepackage{amssymb,bbold}
\usepackage{amssymb} 
\usepackage{mathtools}
\usepackage{amsfonts}
\usepackage[shortlabels]{enumitem}
\usepackage{natbib}
\usepackage[textfont=it]{caption}
\usepackage[textfont=it]{subcaption}
\usepackage{xcolor}
\usepackage{float}

\newcommand{\TT}{\mathbb{T}}

\begin{document}
\title{Clustering Schemes on the Torus\\ with Application to RNA Clashes} 
\titlerunning{Clustering on the Torus}
\thispagestyle{empty}
\author{Henrik Wiechers\inst{1}, Benjamin Eltzner\inst{1}, Stephan F. Huckemann\inst{1}  \and  Kanti V. Mardia\inst{2}}
\authorrunning{Eltzner and Huckemann}
\institute{$^1$ Georg-August-Universit\"at at G\"ottingen, Germany, Felix-Bernstein-Institute for Mathematical Statistics in the Biosciences, \\Acknowledging DFG HU 1575/7, DFG GK 2088, DFG SFB 1465 and the Niedersachsen Vorab of the Volkswagen Foundation\\
$^2$ University of Leeds, UK, Department of Statistics and University  of Oxford, UK,  Department of Statistics\\
}

\maketitle
\thispagestyle{plain}

\begin{abstract}
    Molecular structures of RNA molecules reconstructed from X-ray crystallography frequently contain errors. Motivated by this problem we examine clustering on a torus since RNA shapes can be described by dihedral angles. A previously developed clustering method for torus data involves two tuning parameters and we assess clustering results for different parameter values in relation to the problem of so-called \emph{RNA clashes}.
    This clustering problem is part of the dynamically evolving field of statistics on manifolds. Statistical problems on the torus highlight general challenges for statistics on manifolds. Therefore, the torus PCA and clustering methods we propose make an important contribution to directional statistics and statistics on manifolds in general.

\end{abstract}


\section{Introduction}

\begin{figure}[ht!]
  \centering
  \includegraphics[width=0.9\textwidth]{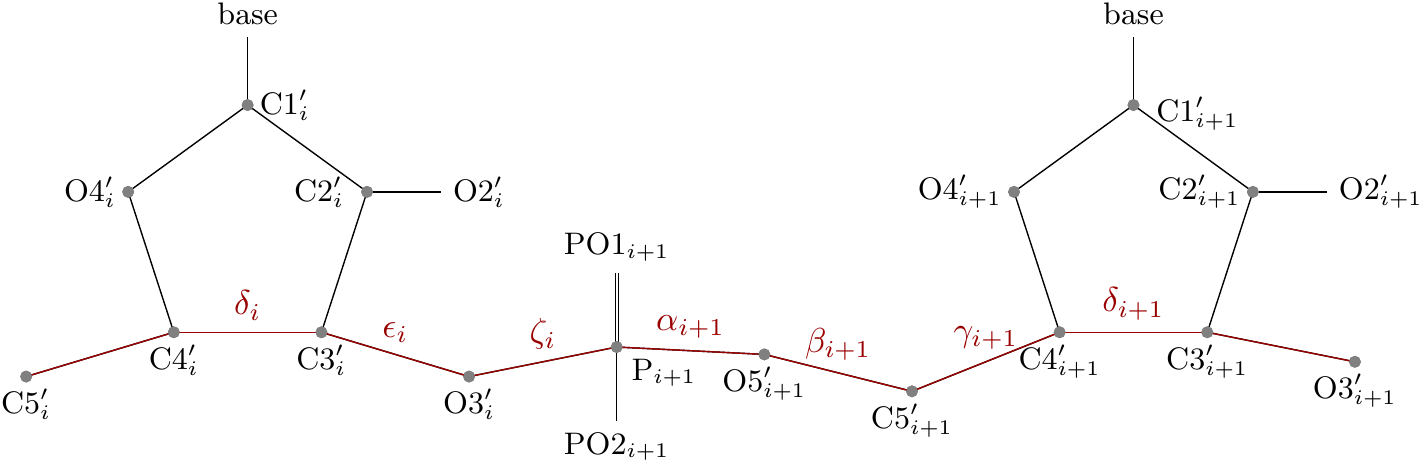}\\
  \vspace*{\baselineskip}
  \includegraphics[width=0.8\textwidth]{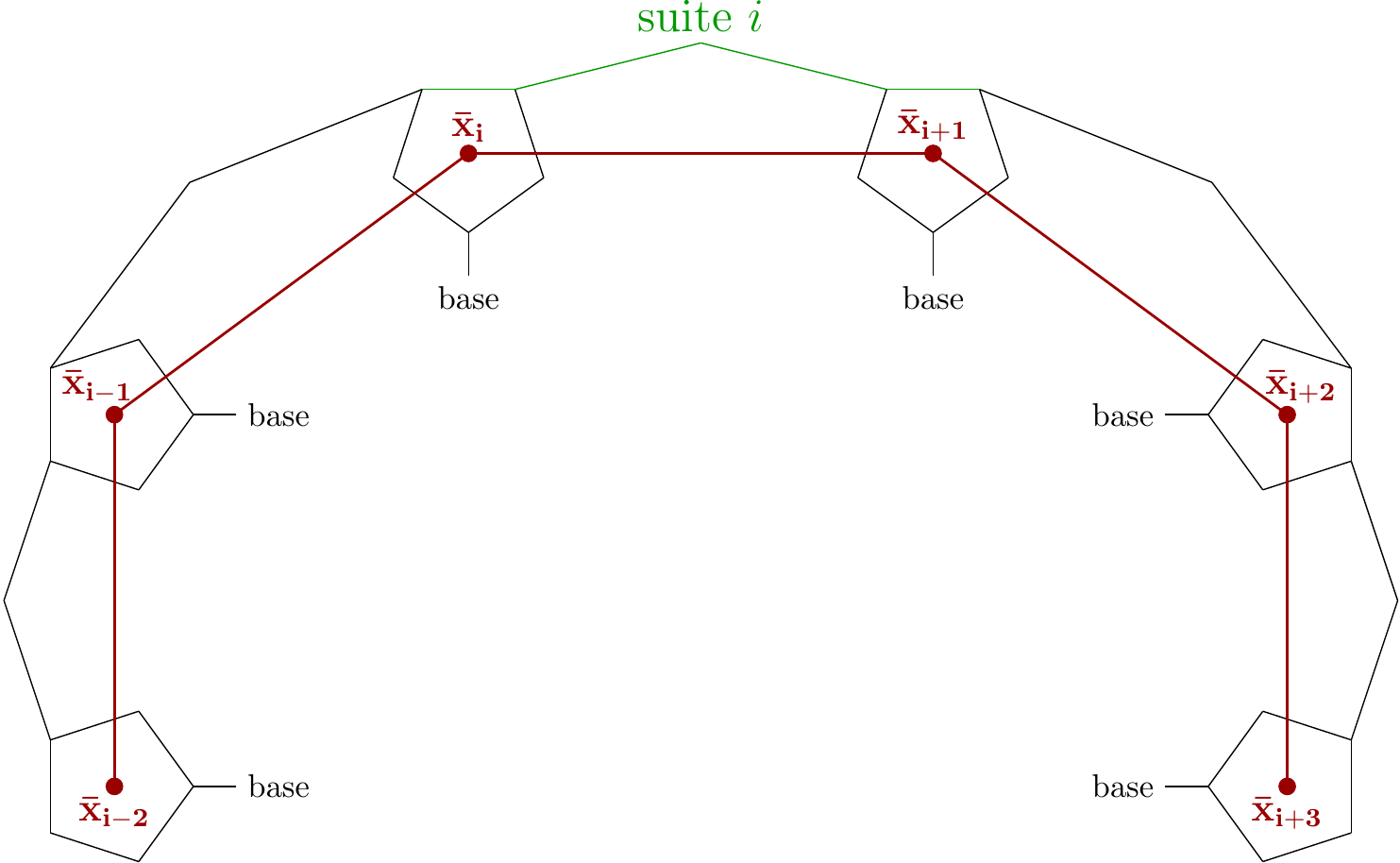}
 \caption{Backbone suite (top) number $i$ with dihedral angles yielding the suite's shape and the centers of the ribose rings (bottom), their connecting backbones (green and black lines) giving 5 suites (two before and two after suite $i$) yielding the mesoscopic shape (red lines). \label{fig:suite-mesoscopic}}
\end{figure}

Inferring \emph{secondary} and \emph{higher order structure} from \emph{primary structure} is one of the holy grails in structural biology.
For \emph{ribonucleotide acid} (RNA), the primary structure is given by a sequence of \emph{nucleobases}. 
RNA 
consists of a \emph{backbone} of alternating \emph{phosphates} and \emph{ribose} sugar rings, to which the nucleobases are attached, see e.g. \cite{eltzner2018} for a brief overview of RNA structure. The part of the backbone from one ribose to the next is called a \emph{suite}, see Figure \ref{fig:suite-mesoscopic} which also gives the different types of atoms.
While to date the primary structure can be quite easily sequenced, see e.g. \cite{Stark2019}, assessing the geometric structure, which is crucial for biological function, requires some effort. Molecular geometry is commonly determined from X-ray crystallography results. However, these reconstructed 3D structures are not error free, but frequently contain \emph{clashes}, see e.g. \cite{Murray13904}.

\begin{definition} 
A \textbf{\emph{clash}} is a forbidden molecular configuration, where two atoms are reconstructed closer to each other than is chemically possible.
\end{definition}
Correction of clashes is essential before data can be fed to learning algorithms, trained on known primary to higher order structure correspondences, which can predict higher order geometric structure from primary structure, see \cite{JAIN2015181}. To avoid ambiguity, we introduce the following terminology.

\begin{definition} 
We consider a single connected RNA strand with $N\geq 6$ consecutive bases indexed in $i \in \{1,\ldots, N\}$, see Figure \ref{fig:suite-mesoscopic}. 
\begin{itemize}
  \item[\textnormal{(1)}] The $i$-th \textbf{\emph{suite}} comprises the RNA region between a $C5_i'$ atom and the second next $O3'_{i+1}$ atom and the backbone shape of the suite is described by the seven dihedral angels $[\delta_{i},\epsilon_i,\dots,\delta_{i+1}]$ for $i = 1,\dots,N-1$.
  \item[\textnormal{(2)}] If two backbone atoms within a suite have a clash with each other, then we call it a \textbf{\emph{clash suite}}.
  \item[\textnormal{(3)}] Each base comes with a pentagonal ribose sugar ring formed by the atoms $\text{C}1'_i$, $\text{C}2'_i$, $\text{C}3'_i$, $\text{C}4'_i$ and  $\text{O}4'_i$. Denoting their center of gravity (i.e. average location) with  $\mathbf{\bar{x}_i}$, for all $i=3,\dots,N-3$, the \textbf{\emph{mesoscopic shape}} corresponding to the $i$-th suite is the similarity size-and-shape in $S\Sigma_3^{6}$ of $X^{(i)}=[\mathbf{\bar{x}_{i-2}}, \mathbf{\bar{x}_{i-1}}, \dots, \mathbf{\bar{x}_{i+3}}]\in \mathbb{R}^{3\times 6}$, see \cite{Drydmard16}.
\end{itemize}
\end{definition}

For the mesoscopic shapes we include the ribose centers of the $k=2$ suites preceding and the $k = 2$ suites following the suite of concern, cf. Figure \ref{fig:suite-mesoscopic}. Choosing $k=2$ reflects the local geometry from a middle viewpoint (mesoscopic) as the $5+1 = 6$ bases from the $2k+1 = 5$ suites correspond roughly to the number of bases involved in a half helix turn, see e.g. \cite{watson2004molecular}.

While there are many types of clashes, most relevant and most difficult to correct are clashes between two backbone atoms (e.g. \cite{Murray13904}) and we deal with such clashes within the same suite.
In contrast to well established methods such as ERRASER from \cite{PMID:23202432} which corrects backbone strands building on elaborate and computationally expensive molecular dynamics (MD) and single nucleotide stepwise assembly (SWA), we determine in \cite{wiechers2021RNA} the nearest neighbors of the \emph{mesoscopic shape} (see Figure \ref{fig:suite-mesoscopic}) corresponding to a clash suite and if a plurality of the suites corresponding to these shapes correspond to a torus cluster, we propose the mean of these suites as corrections. 

Our clash correction approach requires clusters of RNA suites.
RNA suite geometries are very diverse due to two reasons: Firstly, RNA is single-stranded and, secondly, the C2' atom in RNA carries a hydroxyl group. 
Therefore, one can expect many clusters and a high number of outliers; to meet this challenge, we use the \emph{torus clustering} (TC) method from \cite{eltzner2015torus}, which has two tuning parameters $m$ and $d_{\text{max}}$. Here, we investigate typical parameter choices leading to three types of clustering: (a)~\emph{tight}, (b)~\emph{thin} and (c)~\emph{relaxed}.
Tight clustering has been applied for clash correction in \cite{wiechers2021RNA}.

\section{Torus Clustering Correction}
Note that at suite level, a similarity size-and-shape representation would have excessive degrees of freedom because bond lengths and angles are fixed by the laws of chemistry. In consequence, a similarity size-and-shape mean, see e.g. \cite{Drydmard16}, would usually not yield a suite configuration which is possible in reality. Therefore, we describe suite geometry only in terms of the dihedral angles displayed in Figure \ref{fig:suite-mesoscopic}. As detailed in \cite{eltzner2018}, RNA configurations are then represented on the torus $\TT^7 =(\mathbb{S}^1)^7$ with the product distance.

We use the \emph{torus clustering} (TC) method from \cite{eltzner2015torus}, which builds on torus PCA. In contrast to standard PCA methods in Euclidean space, torus PCA is a challenging problem in directional statistics, see e.g. \cite{mardia2009directional} for an introduction into the topic. Significant progress on torus PCA has recently been achieved by \cite{eltzner2018}, see also \cite{eltzner2015dimension,eltzner2017applying}.
In \cite{eltzner2015torus}, suites are pre-clustered based on single linkage clustering depending on two parameters: 
Initially, in the single linkage branch cutting algorithm, see \cite[Appendix A.7]{eltzner2015torus}, branches are cut at distance $d_{max}$ and clusters with less than $m$ points are labeled as outliers. 
In the present paper we subject the resulting clusters to torus PCA from \cite{eltzner2018}, which, highly effectively reducing dimension, usually yields an essentially one-dimensional representation on a circular principal component. On these, we apply circular mode hunting from \cite{eltzner2018}, based on \cite{duembgen2008}, which often leads to a further and final partitioning. 


\section{A Comparison Study}
\label{sec: A Comparison Study}
A classic dataset from the \href{https://www.rcsb.org/}{RCSB PDB Protein Data Bank} is used, which was carefully created by Duarte and Pyle \cite{DUARTE19981465} and extended by \cite{WADLEY2007942} to ensure high experimental X-ray precision (3 \AA{}ngstr\"om). The data set contains 8685 suites. We only work with those suites for which a mesoscopic shape can be defined and we filter out all suites that are involved in clashes between two backbone atoms that are both within the same mesoscopic shape. This leaves 6815 suites, our benchmark data set.

We now compare how different choices for the parameters $d_{\text{max}}$ and $m$ in the pre-clustering affect the resulting overall clustering. Recall from above that due to high variability of RNA structures we expect many clusters and even more outliers.

\subsection{Tight clustering}

\begin{figure}[ht!]
        \centering
        \subcaptionbox{All cluster sizes of tight clustering. Asterisks mark clusters displayed in panel~(b).
        \label{tab: cluster sizes clustering 1}}[0.36\textwidth]{
        \begin{tabular}{c|c}
            Cluster number & Size\\ 
            \hline
            1$^*$ &  4382 \\
            2~ & 477 \\
            3~ & 223 \\
            4~ & 107 \\
            5$^*$ & 92 \\
            6~ & 56 \\
            7$^*$ & 43 \\
            8$^*$ & 40 \\
            9~ & 31 \\
            10$^*$ & 27 \\
            11~ & 26 \\
            12~ & 22 \\
            Outliers & 1289\\
            \hline
            Total & 6815
        \end{tabular}
        }
    \hspace*{0.02\textwidth}
     \subcaptionbox{Five exemplary clusters that can be well displayed together from tight clustering. Suites are mean aligned for illustration purposes using the generalised Procrustes algorithm.}[0.55\textwidth]{
        \includegraphics[width=0.55\textwidth]{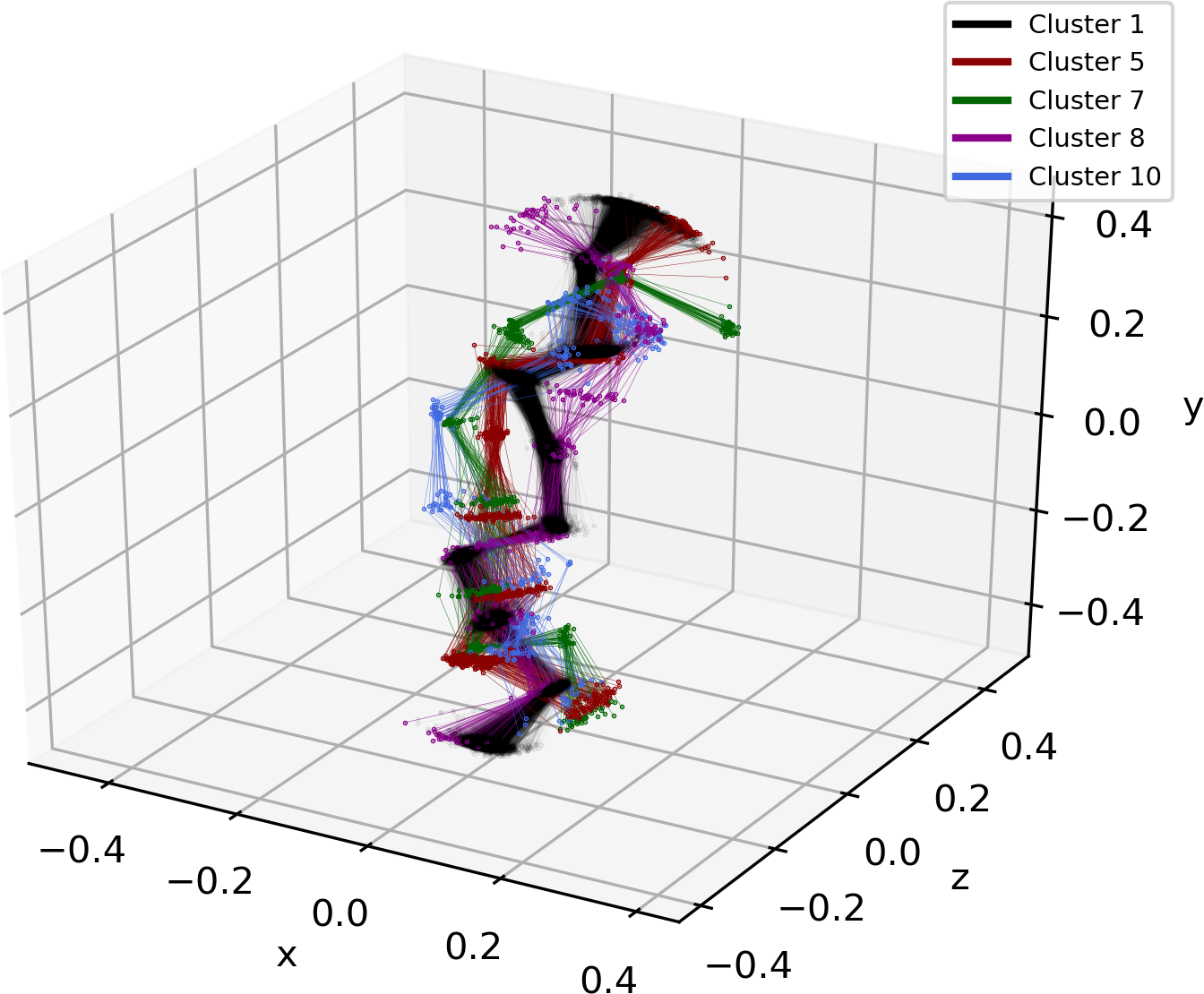}
    }
    \caption{Tight clustering: all clusters and outliers (left), five selected clusters (right). \label{fig:tight-clust}}
\end{figure}

In  \emph{tight clustering} we aim at larger and concentrated clusters at the price of a larger number of outliers which cannot be allocated to any of the clusters. We choose $m=20$ and $d_{\text{max}}$ such that $12.5\%$ of the suites in the single linkage tree are in a branch with less than $m$ data points. 
The clustering results are summarized in 
Figure \ref{tab: cluster sizes clustering 1}.

This clustering returned 12 clusters, the largest corresponding to the A helix shape contains 4382 elements is highly dominant. All clusters are rather dense and even the smallest cluster has a credible size with 22 elements. The number of outliers (1289), however, is quite large.


\subsection{Thin clustering}

\begin{figure}[ht!]
        \centering
        \subcaptionbox{All clusters from thin clustering
        \label{tab: cluster sizes clustering 2}}[0.36\textwidth]{
        \begin{tabular}{c|c}
            Cluster number &  Size  \\
            \hline
            1~ &  4382 \\
            2~ & 527 \\
            3~ & 223 \\
            4~ & 163 \\
            5~ & 141 \\
            6~ & 132 \\
            7~ & 92 \\
            8$^*$ & 78 \\
            9~ & 75 \\
            10~ & 72 \\
            11~ & 57 \\
            12~ & 11 \\
            13~ & 46 \\
            14~ & 45 \\
            15~ & 36 \\
            16~ & 20 \\
            17~ & 20 \\
            Outliers & 659\\
           \hline
            Total & 6815
        \end{tabular}
        }
    \hspace*{0.02\textwidth}
    \subcaptionbox{The eighth cluster from thin clustering illustrates that clusters with high variance appear in the clustering. The suites are mean aligned as in Figure \ref{fig:tight-clust}.\label{fig: cluster sizes clustering 2}}[0.6\textwidth]{
    \includegraphics[width=0.6\textwidth]{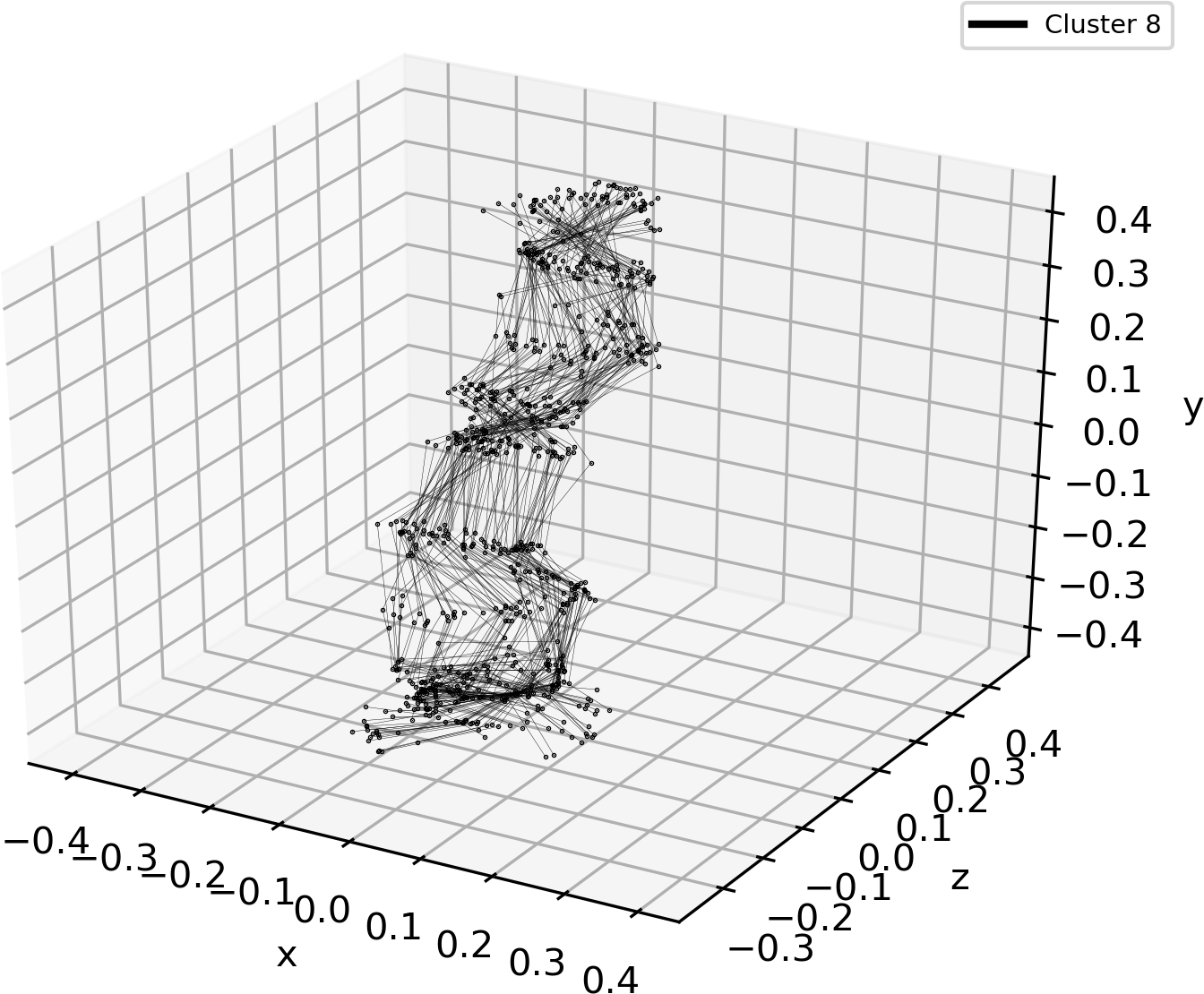}
    }
    \caption{Thin clustering: all clusters and outliers (left), selected cluster of thin density (right). 
    \label{fig:relaxed-clust}}
\end{figure}

In \emph{thin} clustering we aim at reducing the number of outliers while keeping clusters sizes credible. Thus we choose $m=20$ as before, but change $d_{\text{max}}$ so that only $5\%$ of the suites in the single linkage tree are in a branch with less than 20 data points. Consequently, we have the same minimal cluster size, but a larger variance within the clusters and fewer outliers, as 
summarized in Figure \ref{tab: cluster sizes clustering 2}.

The first cluster corresponds to the first cluster in tight clustering and has the same number of elements. The other clusters in tight clustering also occur in thin clustering, but have more elements in thin clustering. As illustrated in Figure \ref{fig: cluster sizes clustering 2}, there are some clusters with large variance. The reason for this is that single linkage clustering often results in thin and elongated clusters due to chaining effects. This effect is more pronounced the lower the number of outliers in the clustering, which is mainly determined by $d_{\text{max}}$.

\subsection{Relaxed clustering}
\label{seq: multi mode clustering}
\begin{figure}[ht!]
        \centering
        \subcaptionbox{All clusters from relaxed clustering
        \label{tab: cluster sizes clustering 3}}[0.36\textwidth]{
        \begin{tabular}{c|c}
            Cluster number & Size  \\
            \hline
            1~ &  4382 \\
            2~ & 514 \\
            3~ & 217 \\
            4~ & 103 \\
            5~ & 98 \\
            6~ & 69 \\
            7~ & 64 \\
            8~ & 53 \\
            9~ & 45 \\
            10~ & 43 \\
            11~ & 43 \\
            12$^*$ & 40 \\
            13~ & 29 \\
            14~ & 27 \\
            15~ & 23 \\
            16~ & 22 \\
            Outliers & 1043\\
           \hline
            Total & 6815
        \end{tabular}
        }
    \hspace*{0.02\textwidth}
    \subcaptionbox{The 12th cluster from relaxed clustering illustrates that some clusters appear that visually could
    be decomposed into multiple sub-clusters or modes. The suites are mean aligned as in Figure \ref{fig:tight-clust}. 
    \label{fig: cluster sizes clustering 3}}[0.6\textwidth]{
    \includegraphics[width=0.6\textwidth]{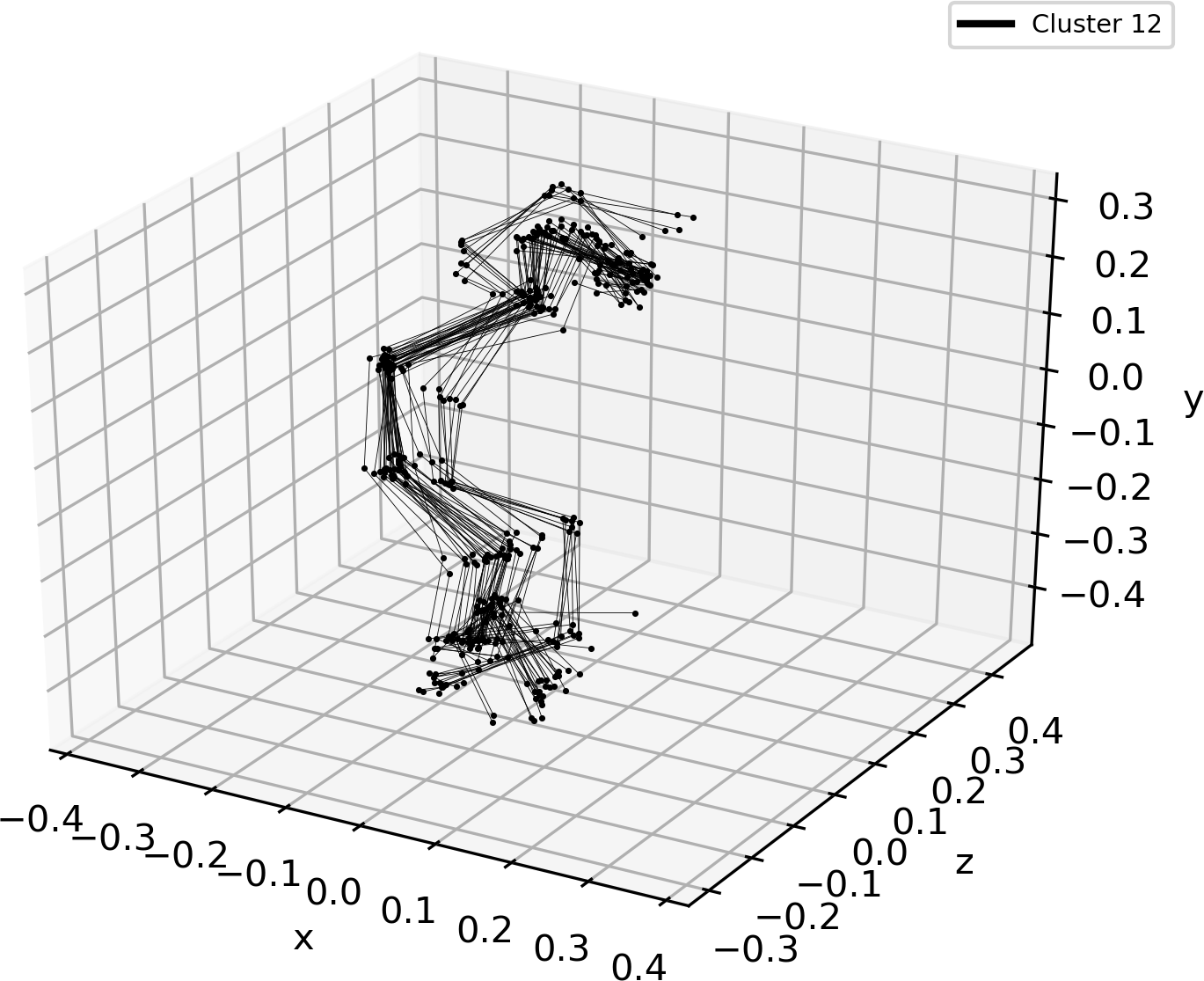}
    }
    \caption{Relaxed clustering: all clusters and outliers (left), selected cluster of relaxed density (right).
    }
\end{figure}

In \emph{relaxed clustering} we relax the minimal cluster size, here to $m=5$, but choose $d_{\text{max}}$ such that $12.5\%$ of the suites in the single linkage tree are in a branch with less than $m$ data points, as in tight clustering.
The clustering results are summarized in Figure \ref{tab: cluster sizes clustering 3}.
Again, the first cluster corresponds to the first cluster in tight clustering and has the same number of elements. The other clusters from the tight clustering also occur in relaxed clustering, but have different sizes. Remarkably, some clusters, such as the one in Figure \ref{fig: cluster sizes clustering 3}, have visually more than one mode. The reason for this is that in single linkage branch cutting, which is used as pre-clustering, several different clusters having at least $m$ elements are sometimes interpreted as one cluster. For small $m$ such additional modes may contain only very few elements, so that the power of the multiscale test by \cite{duembgen2008}, translated to the circular case by \cite{eltzner2018}, does not suffice to separate them.


\section{Discussion}

The task to correct clashes from reconstructions of biomolecules is currently met with high interest in statistics and biophysical chemistry.
As powerful statistical learning methods are based on clustering we investigated the challenge of \emph{torus clustering} (TC), i.e. clustering on the torus.
Specifically we have clustered geometric diversity of  non-clashing reconstructions of biomolecules with the aim of  corrected clashing reconstructions by suitably assigning to clusters in \cite{wiechers2021RNA}.
Our TC correction method hinges on two parameters, \emph{minimal cluster size} $m$ and  \emph{maximal outlier distance} $d_{\text{max}}$ and in Section \ref{sec: A Comparison Study} we have reported outcomes for typical parameter choices. These represent many more experiments, we have conducted. 

Since the highly efficient underlying torus PCA yields mainly one-dimensional pre-clusters, we can apply circular mode hunting, and in order to build on statistical significance, we find $m \approx 20$ rather reasonable. For smaller $m$ such as $m=5$ yielding relaxed clustering from Section \ref{seq: multi mode clustering}, several modes within clusters are visible, they cannot be separated, however, with statistical significance.

Control of outliers is closely related to cluster density, it turns out that choosing $d_{\text{max}}$ resulting in about $1$ outlier out of $8$  yields clusters of convincing shape. Aiming at less outliers as in thin clustering, cluster are less tight and new thin clusters appear (see Figure \ref{fig: cluster sizes clustering 2}), which are of poorer quality.

As a result of this study,  we sum up that tight clustering appears to be the most promising for the task of clash correction. 
Even though statistical clustering is in principle a method of unsupervised learning, the pursuit of optimal choices of tuning parameters warrants supervised learning. Creating suitable objective functions involving expert knowledge remains  a challenge.


\end{document}